\documentstyle{article}

\begin{document}

\rightline{SAGA-HE-84-95}

\rightline{June 1995}

    \centerline{{\bf Nucleon mean free path in nuclear matter based on
                  nuclear Schwinger-Dyson formalism }}

    \centerline{Tomohiro MITSUMORI, Nobuo NODA,Kazuharu KOIDE }

    \centerline{, Hiroaki KOUNO , Akira HASEGAWA}

    \centerline{{\it Department of Physics, Saga University, Saga 840,
    Japan}}

    \centerline{~and}

    \centerline{Masahiro NAKANO}

    \centerline{{\it University of Occupational and Environmental
     Health}}

    \centerline{{\it Kitakyushu 807, Japan}}

{}~~~~~~~~~~~

{}~~~~~~~~~~~~~
{\bf abstract}
A  mean free path of nucleon moving through  nuclear matter
 with kinetic energy of more than 100MeV is formulated based on the bare
vertex nuclear Schwinger-Dyson (BNSD) method in the
 Walecka model. The self-energy which is derived from the higher
order diagrams more than the forth order includes the Feynman part of
propagator of energetic nucleon and grows up rapidly as an increase of
 kinetic energy.
 To avoid too large growth of these diagrams,  meson propagators are
 modified by introducing some form factors
 to take account of a internal structure of hadron.
 It is confirmed that the mean
 free path calculated by the BNSD method agrees good with experimental data
 if a reasonable form factor is
 chosen, i.e., a dipole (quadrupole) type
 of form factor with a cut-off parameter about 750 MeV $\sim$ 1000 MeV
 (1200 MeV $\sim$ 1500 MeV).

\vfill\eject

{}~~~~~~~~~~~~~~~~~~~~

{\bf 1. INTRODUCTION}

{}~~~~~~~~~~~~~~~~~~~~

The nucleon mean free path in nuclear medium is one of the important
quantity to show the property of nuclear matter as well as the binding energy
 and the incompressibility etc.
 It is experimentally extracted from the proton (or neutron)-nucleus total
 cross sections for many targets at a spread of energy[1][2].

Recently there have been the numerical results of the nuclear
mean free path in nuclear matter
in the framework of relativistic approach based on the Dirac
phenomenology [3] , in the relativistic impulse approximation [4][5] and
in the density dependent Hartree-Fock approximation [6].In this paper
 we evaluate the nucleon mean free path in nuclear matter in the $\sigma$-
$\omega$ model based on the nuclear Schwinger-Dyson formalism .

In the last several years ,the two loop correction and the ring-sum
correction were
calculated in the framework of renormalizable $\sigma -\omega$ model
 and these contributions to the energy density are much in the Feynman part
and in the finite density part [7][8][9][10]. It is known that the large
 contributions of the Feynman part
are reduced by introducing form factors at all vertex
 in the loop diagram
 [11][12] or by vertex corrections[13]. These procedures for reducing
short-range distance contributions are qualitatively acceptable at present
 to take account of the internal structure of hadron.
By using the bare vertex nuclear Schwinger-Dyson (BNSD) method we showed
  that the loop
 corrections  in the density part were also hardly reduced
  because the these corrections
 were strongly canceled among the contributions of three components
,i.e.,   $\sigma$- and $\omega$-mesons and $\sigma$-$\omega$ mixture
 [14][15].

  Until now we have studied about the saturation property of ground state
  of nuclear matter
   based on the BNSD method[14][15]. In these studies,
   we determined two meson-nucleon
  coupling constants to get the minimum energy at the normal density,
 taking care to
    remove the instability near the normal density by using some
   recipes [16][17]. After this work, we studied
    optical potentials for nucleon near Fermi surface under the BNSD
    approximation
    by using the results of
    nuclear matter[17], and
    we obtained the  reasonable
     optical potential both in the real and imaginary parts
     compared with empirical findings
      [18][19].

       In this paper we estimate the mean free path of nucleon which
  travels through nuclear matter having the kinetic energy of more than
  100 MeV above the Fermi surface. We will show that an optical potential for
 such a high energy nucleon outside the Fermi sphere is derived from the
Feynman part of self-energy which corresponds to the higher order
diagrams more
 than the fourth-order, besides the density part of self-energy. These
 higher order diagrams yield very large contributions to the optical potential
 both in the real and imaginary parts as well as large contributions of the
 Feynman part of energy density of nuclear matter. We point out that there
 need  meson propagators modified by introducing some form factors to make
 the mean free path calculated by the BNSD method agree with ones
 extracted  in this kinetic energy region.

 The organization of this paper is as follows. We develop the
 BNSD method for a energetic
nucleon traveling far above the Fermi surface
  in Sec. II.
The numerical results and discussion are shown in Sec. III.
We summarize our work of this paper in Sec.IV.

{}~~~~~~~~~~~~~~~~~~~~

{\bf 2. FORMULATION}

{}~~~~~~~~~~~~~~~~~~~~

We adopt the Walecka model[20] which consists of three fields, the
nucleon$\psi$,the scalar $\sigma$-meson $\phi$ and the vector
$\omega$-meson
$V_\mu$.The lagrangian density is given by

$$L = -\bar\psi(\gamma_\mu\partial_\mu +M)\psi -{1\over 2}(\partial_\mu
\phi\partial_\mu\phi +m_s^2 \phi^2) $$

$$  -({1 \over 4}F_{\mu\nu}F_{\mu\nu}+{1 \over 2}m_v^2 V_\mu V_\mu)
+g_s\bar\psi\psi\phi +ig_v\bar\psi\gamma_\mu\psi V_\mu  , \eqno(1) $$

\noindent
where $F_{\mu\nu}= \partial_\mu V_\nu - \partial_\nu V_\mu$ and $M$, $m_s$,
 $m_v$, $g_s$ and $g_v$ are nucleon mass, $\sigma$-meson mass,
$\omega$-meson mass, $\sigma$-nucleon and $\omega$-nucleon
coupling-constants respectively.

The nucleon propagator is obtained by following form,

$$G(k) = G_F(k)+G_D(k) $$

$$= {-1 \over i\gamma_\mu k^*_\mu +M^*_k -i\epsilon}
      +(-i\gamma_\mu k_\mu^* +M_k^* ){i\pi \over E^*_k}
\delta (k_0^* -E_k^* )\theta (k_F -|{\bf k}|),\eqno(2) $$

$$E_k^* =\sqrt{{\bf k}^{*2}+M_k^{*2}} , \eqno(3)$$

$$M_k^* =M+\Sigma_s(k) , \eqno(4)$$

$$k_\mu^* = ( {\bf k}^* , ik_0^*  ) $$

$$=\biggr ( {\bf k} (1+\Sigma_v (k)),i(k_0 +\Sigma_0 (k)) \biggr )
              , \eqno(5)$$

\noindent
where $G_F$ and $G_D$ are the Feynman part and
the density-dependent part, respectively.

The nucleon self-energy $\Sigma (k)$ is classified as shown in Fig.1

\centerline{$\underline{\overline{\rm Fig.1}}$}

$$\Sigma (k) = \Sigma_H (k) + \Sigma_{SDD}(k) + \Sigma_{SDF}(k) ,\eqno(6)$$

\noindent
where $\Sigma_H ,\Sigma_{SDD}$ and $\Sigma_{SDF}$ are the Hartree
term ,the SD density term and the SD Feynman term respectively.
And this nucleon self-energy is alternatively classified according to
the property of Lorentz transformation,

$$\Sigma = \Sigma_s (k)-\gamma_0 \Sigma_0 (k) +i  \gamma_i \cdot
k_i \Sigma_v (k), \eqno(7)$$

\noindent
where $\Sigma_s ,\Sigma_0$ and $\Sigma_v$ mean the scalar component
,the timelike component of vector type and the spacelike component
of vector type.

In the BNSD method, the meson propagator satisfies the following
Dyson equation as shown in Fig.1,

$$D(q)=D_0 (q)+D_0(q)\Pi (q)D(q) ,   \eqno(8)$$

\noindent
where $D_0(q)$ denotes the free meson propagator and $\Pi (q)$
denotes the
 meson self-energy.
Then the meson propagator has the real and the imaginary parts,

$$D(q) = Re D(q)+iIm D(q) . \eqno(9)$$

\noindent
To derive the imaginary part of self-energy we execute a Wick rotation
$q_0$ $\rightarrow$ $iq_0$ in the $q_0$-integral of the SD Feynman term
 and rewrite $\Sigma_{SDF}$ as follows,

$$\Sigma_{SDF} = \Sigma_{SDF}^{on} + \Sigma_{SDF}^{off} ,
              \eqno(10)$$

$$Re\Sigma_{SDF} = Re\Sigma_{SDF}^{on} + Re\Sigma_{SDF}^{off} ,
              \eqno(11)$$

$$Im\Sigma_{SDF} = Im\Sigma_{SDF}^{on}  .
              \eqno(12)$$

\noindent
The superscripts "on" or "off" of $\Sigma_{SDF}$ denote that
 the nucleon propagator included in the SD diagram is on-shell or
 off-shell, respectively. We note that the real part of $\Sigma_{SDF}$
 is composed of two terms, i.e., a finite term Re$\Sigma_{SDF}^{on}$
 and an infinite term Re$\Sigma_{SDF}^{off}$, while the imaginary
 part of $\Sigma_{SDF}$ is only one term, a finite
 Im$\Sigma_{SDF}^{on}$. We drops the infinite real self-energy as
 stated in the previous paper [17].

\noindent
We obtain the explicit expression of real
and imaginary nucleon self-energies corresponding to the Feynman diagrams
 given in Fig.2 as follows,

\centerline{$\underline{\overline{\rm Fig.2}}$}

$$Re \Sigma_s (k) = -{2 \over \pi^2}{g_s^2 \over m_s^2}\rho_s-{g_s^2 \over
8\pi^2}\int_{0}^{k_F}  q^2 {M^*_q \over E^*_q} dq \int_{-1}^{1}dx  \Delta_0(R)
$$

$$ + {g_v^2 \over 8\pi^2} \int_{0}^{k_F}  q^2{M^*_q \over E^*_q} dq
\int_{-1}^{1}  dx  4 D_0 (R) $$

$$ + {g_s^2 \over 8\pi^2} \int_{k_F}^{k}  q^2 dq  \int_{-1}^{1}  dx
{M^*_q \over E^*_q} [{R^2 \over \vec R^2} (\Delta_0 \Pi_m D_m)_R
+(\Delta_0 \Pi_s D_s)_R] $$

$$ + {g_v^2 \over 8\pi^2} \int_{k_F}^{k}  q^2 dq  \int_{-1}^{1}  dx
{M^*_q \over E^*_q} [ -4(D_0 \Pi_t D_t)_R +{ R^2 \over \vec R^2}\{
 -(D_0 \Pi_l D_l)_R +(D_0 \Pi_t D_t)_R -{ R^2 \over \vec R^2}
(D_0 \Pi_m D_m)_R \} ] $$

$$ + {2 g_s g_v \over 8\pi^2}\int_{k_F}^{k} q^2 \int_{-1}^{1}dx
{ R^2 \over \vec R^2} Re D_m(R), \eqno(13a) $$

$$Im \Sigma_s (k) = {g_s^2 \over 8\pi^2}\int_{k_F}^{k}  q^2 {M^*_q \over E^*_q}
dq \int_{-1}^{1}dx ImD_s(R) $$

$$ - {g_v^2 \over 8\pi^2} \int_{k_F}^{k}  q^2{M^*_q \over E^*_q} dq
\int_{-1}^{1}  dx \{ 4ImD_l (R)+ (3+ {R_0^2 \over \vec R^2} )(ImD_t (R)-ImD_l
(R)) \} $$

$$ + {g_s g_v \over 4\pi^2}\int_{k_F}^{k} q^2 dq  \int_{-1}^{1}dx ImD_m (R) {
R^2 \over \vec R^2}, \eqno(13b)$$

$$Re \Sigma_0 (k) = -{2 \over \pi^2}{g_v^2 \over m_v^2}{k_F^3 \over 3}+{g_s^2
\over 8\pi^2} \int_{0}^{k_F} q^2 dq \int_{-1}^{1}dx \Delta_0(R) $$

$$ + {g_v^2 \over 8\pi^2} \int_{0}^{k_F} q^2 dq \int_{-1}^{1} dx
 2 D_0 (R) $$

$$ - {g_s^2 \over 8\pi^2} \int_{k_F}^{k}  q^2 dq  \int_{-1}^{1}  dx
 [{R^2 \over \vec R^2} (\Delta_0 \Pi_m D_m)_R
+(\Delta_0 \Pi_s D_s)_R] $$

$$ + {g_v^2 \over 8\pi^2} \int_{k_F}^{k}  q^2 dq  \int_{-1}^{1}  dx
 [ -2(D_0 \Pi_t D_t)_R -{ R^2 \over \vec R^2}\{
 -(D_0 \Pi_l D_l)_R +(D_0 \Pi_t D_t)_R -{ R^2 \over \vec R^2}
(D_0 \Pi_m D_m)_R \} ] $$

$$ - {2 g_s g_v \over 8\pi^2}\int_{k_F}^{k} q^2 \int_{-1}^{1}dx
{M^*_q \over E^*_q}{ R^2 \over \vec R^2} Re D_m(R), \eqno(14a)$$

$$Im \Sigma_0 (k) = -{g_s^2 \over 8\pi^2} \int_{k_F}^{k} q^2 dq \int_{-1}^{1}dx
ImD_s(R) $$

$$ - {g_v^2 \over 8\pi^2} \int_{k_F}^{k} q^2 dq \int_{-1}^{1} dx \{ 2ImD_l
(R)+(2+{R^2 \over \vec R^2})(ImD_t (R)-ImD_l (R))\} $$

$$ - {g_s g_v \over 4\pi^2}\int_{k_F}^{k} q^2 dq  \int_{-1}^{1}dx  {M_q^* \over
E_q^*}{ R^2 \over \vec R^2} ImD_m (R) , \eqno(14b) $$

$$Re \Sigma_v (k) = {g_s^2 \over 8\pi^2 k^2} \int_{0}^{k_F} q^2 dq
\int_{-1}^{1}dx
 {q^* kx \over E_q^*} \Delta_0(R) $$

$$ + {g_v^2 \over 8\pi^2  k^2} \int_{0}^{k_F} q^2 dq \int_{-1}^{1} dx
{q^* k x \over E_q^*} 2D_0 (R) $$

$$ - {g_s^2 \over 8\pi^2 k^2} \int_{k_F}^{k}  q^2 dq  \int_{-1}^{1}  dx
{q^* k x \over E_q^*}  [{R^2 \over \vec R^2} (\Delta_0 \Pi_m D_m)_R
+(\Delta_0 \Pi_s D_s)_R] $$

$$ + {g_v^2 \over 8\pi^2 k^2} \int_{k_F}^{k}  q^2 dq  \int_{-1}^{1}  dx
 {q^* k x \over E_q^*} [ -2(D_0 \Pi_t D_t)_R -{ R^2 \over \vec R^2}\{
 -(D_0 \Pi_l D_l)_R +(D_0 \Pi_t D_t)_R -{ R^2 \over \vec R^2}
(D_0 \Pi_m D_m)_R \} ] \eqno(15a)$$

$$Im \Sigma_v (k) = -{g_s^2 \over 8\pi^2 k^2} \int_{k_F}^{k} q^2
  dq \int_{-1}^{1}dx { q^* kx \over E^*_q} ImD_s(R) $$

$$ - {g_v^2 \over 8\pi^2  k^2} \int_{k_F}^{k} q^2 dq \int_{-1}^{1} dx  {q^* k x
\over E_q^*}\{ 2ImD_l (R) +(2-{R^2 \over \vec R^2})(ImD_t (R)-ImD_l (R))\},
\eqno(15b)$$

\noindent
where $R=k-q$, and $\rho_s$ and $\rho_B$ denote the scalar
and the baryon densities respectively, and $\Delta_0$ and $D_0$ denote the
 free propagators of $\sigma$- and $\omega$-mesons, respectively.
 The subscripts s,l,t and m of $D(R)$ and $\Pi (R)$ denote the component of
 $\sigma$-meson, the longitudinal and transverse components of
 $\omega$-meson and the component of mixture of $\sigma$- and
 $\omega$-mesons,
respectively.
The detailed expressions of $D(R)$ are given in Ref.[16] and
the analytical expressions of $\Pi (R)$ are given in Ref.[10].

The last three terms in Eqs.(13a) and (14a) and the last two terms in Eq.(15a)
 are contributed from the same higher order
diagrams which yield the imaginary part.

We modify the meson
propagators in the SD Feynman part of self-energy
by introducing some form factors
at the vertices in agreement with the point of view in Ref.
[11][12][21] to examine the short range
interactions and the finite size of hadrons ,as follows

$$D (q^2) \rightarrow [F(q^2)]^2 \cdot D(q^2), \eqno(16)$$

\noindent
where $F(q^2 )=1/[1+q^2 / \Lambda^2 ]^n$ . The case of n=1 is a
monopole type of
form factor and n=2 is a dipole type, etc.

Optical potentials are defined from self-energies $\Sigma$ as

$$U_S = {\Sigma_s - M\Sigma_v  \over 1+\Sigma_v} = U_{SR} +iU_{SI},
\eqno(17) $$

$$  U_V = {-\Sigma_0 + E\Sigma_v \over 1+\Sigma_v} =
U_{VR} +iU_{VI}, \eqno(18) $$

\noindent
where $E$ is the energy of objective nucleon propagating with the
momentum $k$
 satisfying the following dispersion relation,

$$E =\sqrt{ {\bf k}^2 +(M+U_S )^2 }+U_V . \eqno(19)$$

\noindent
Eq.(19) is rewritten as

$${ {\bf k}^2 \over 2M} +V+iW =E-M+{ (E-M)^2 \over 2M}, \eqno(20)$$

\noindent
with Schr$\ddot {\rm o}$dinger equivalent potential form

$$V = U_{SR}+U_{VR}+{ (E-M) \over M}U_{VR}+{1 \over 2M} (U_{SR}^2
+U_{VI}^2-U_{SI}^2-U_{VR}^2 ), \eqno(21) $$

$$W = U_{SI}+U_{VI}+{ (E-M) \over M}U_{VI}+{1 \over M}
(U_{SR}U_{SI}-U_{VR}U_{VI} ). \eqno(22) $$

\noindent
The nucleon momentum is complex as well as the optical potentials and
can be expressed as

$$ |{\bf k}| = k_R +ik_I . \eqno(23)$$

\noindent
and then we obtain the nucleon mean free path as follows,

$$\lambda ={1 \over 2k_I}$$

$$={1 \over 2}\Biggr \{ -M \biggr (E-M-V +{ (E-M)^2 \over 2M }\biggr )+M
\biggr[ \biggr (E-M-V +{ (E-M)^2 \over 2M }\biggr )^2 +W^2\biggr ]^{1/2} \Biggr
\}^{-1/2}. \eqno(24)$$

{}~~~~~~~~~~~~~~~~~~~~

{\bf 3. RESULTS AND DISCUSSION}

{}~~~~~~~~~~~~~~~~~~~~

In this section we calculate self-energies  of the nucleon
 propagating with the energy $E$ and the momentum ${\bf k}$ ,
convert them into the  optical potentials
in the Schr$\ddot {\rm o}$dinger equivalent form , and evaluate the
 nucleon mean free path from Eq.(24). The relation between
 the energy $E$ and the  momentum $k$ is given by approximating
 Eq.(19) as follows,

$$E=\sqrt{{\bf k}^2  +(M+Re\Sigma_s (E,k))^2 }
    -Re\Sigma_0 (E,k) .\eqno(25)$$

  \noindent
  The nucleon momentum ${\bf k}$, the upper limit of integral in the real and
 imaginary
  self-energy, is determined
  from  Eq.(25) by putting the nucleon energy $E=E_{in} +M$
  ,where $E_{in}$ is the nucleon incident energy.

In this paper we determine the coupling constants of the $\sigma$-nucleon and
$\omega$-nucleon
 to satisfy the saturation property of nuclear matter at
 the normal density $\rho$=0.170 fm$^{-3}$ ($k_F$ = 1.36 fm$^{-1}$)
 , and 0.193 $fm^{-3}$ ($k_F$ = 1.42 $fm^{-1}$) and summarize the parameters in
 Table.

\centerline{$\underline{\overline{\rm Table}}$}

Using these parameters (we choose the parameters in the case of
the normal density $\rho$=0.17 fm$^{-3}$ ),
 we calculate the nucleon mean free path.
We start with evaluating the higher order diagrams more  than the fourth
 order in the nucleon self-energy shown in Fig.2.
 Since these diagrams
  include the Feynman propagator of nucleon on-shell in the intermediate
  state,
   we are afraid whether these diagrams give
   a large contribution to the self-energy.
  In Fig.3, we show contributions of  $\sigma$-meson, $\omega$-meson
  and $\sigma$-$\omega$ mixture to Re$\Sigma_s$ and Re$\Sigma_0$,
  respectively. As the increase of kinetic energy of nucleon, $E-M$,
  each component, $\Sigma^\sigma$, $\Sigma^\omega$ and
  $\Sigma^{\sigma-\omega}$ , grows almost linearly, because the meson
  propagator $\sim 1/q^2$ and the Jacobian $q^2$ factor
  cancel each other in $q$-integrals
 of self-energies with the upper limit of $k$.
 In Re$\Sigma_s$, additive contributions of $\sigma$-meson and
 $\sigma$-$\omega$ mixture cancel strongly with the
  one of
 $\omega$-meson and as a result the net contribution is small.
  On the
 other hand, contributions of three components to Re$\Sigma_0$ are
 all additive and as a result the self-energy is very huge
 and the net contribution
  becomes nearly equal with the Hartree contribution.

 \centerline{$\underline{\overline{\rm Fig.3(a) , Fig.3(b) }}$}

In Fig.4, we also show the contributions of three components to
 Im$\Sigma_s$
 and Im$\Sigma_0$, respectively.
 The contributions of
 $\sigma$-meson and $\sigma$-$\omega$ mixture to Im$\Sigma_0$
 are far smaller than
 the one of $\omega$-meson.
 The
 $\omega$-meson dominance should be marked. The value of Im$\Sigma_0$
 at $E-M = 200 MeV$ are considerably reduced in comparison with the one
  derived from the
 fourth order diagrams in Ref.[21].
 The higher order contribution more than the fourth order is very important
[17].
  The value of Im$\Sigma_s$
  is also reduced  at the
 same kinetic energy and, however, has the negative sign
  when $E-M$ $>$ $350 MeV$,  which is
 opposite to the sign of Im$\Sigma_s$ obtained from the fourth order
 diagram.
 The change of sign is easily understood from Fig.4(b).
  In lower kinetic energy region Im$\Sigma_s$ is positive because
 the contribution of $\omega$ meson is large and positive
 compared with the small and negative contributions of others.
 On the other hand, in higher kinetic energy region, the sign of
 Im$\Sigma_s$ changes to the negative one because the contribution of
 $\omega$ meson changes to the negative one and becomes  large.
 It should be remarked that Im$\Sigma$ originates from the Feynman
 part of self-energy. So, although Im$\Sigma$ is small if
 $E-M$ $<$ $100 MeV$,
  it grows up unphysically
 if $E-M$ $>$ $100 MeV$ as well as the  vacuum effect
 for the energy density in nuclear matter
 is unphysically large. Then, we introduce form
 factors into each vertex by modifying meson propagators,
  $D(q) \rightarrow [f(q^2)]^2\cdot D(q)$ as shown in Ref.[21].
 It is noted that, since the SD diagrams are composed of full meson propagators
with
 a ring-sum correction, a dipole (quadrupole) type of form factor
 introduced into the SD diagram corresponds to a monopole (dipole) type
 of form factor into the fourth order diagram.

 \centerline{$\underline{\overline{\rm Fig.4(a) , Fig.4(b) }}$}

There are two kinds of $\Sigma_{SDF}$ leading to the  real
self-energy.
One is the above-discussed self-energy derived from the same diagram
 which yields the imaginary part. The other is the self-energy derived
  from the
 diagrams including the Feynman propagator of nucleon off-shell.
 The renormalization procedure showed that Re$\Sigma_{SDF}$ is
  very large.
  So we must develop
  a new recipe which takes account of the size of hadron.
   Then, we should evaluate the two diagrams at the same
 time when we want to know their contributions to the real
 self-energy. In the present situation, therefore, we do not pick
 up both of them though we are afraid of the violation of the
 dispersion relation between the real part and the imaginary part.

 In Fig.5(a) and (b), we show the imaginary part and the real part of
 Schr$\ddot {\rm o}$dinger equivalent potentials
 , in cases of two types of form factor, the dipole with $\Lambda$
 = 750 (solid curve), 1000 (dotted curve) and 1500 MeV (dashed curve)
 and the quadrupole with $\Lambda$ = 1200 (bold solid curve) and
 1500 MeV (bold dotted curve). The solid dots are empirical information
 [18][19].
 As the kinetic energy increases, the imaginary potential decreases slowly
 if $E-M$ $<$ $300 MeV$ and rapidly if $E-M$ $>$ $350 MeV$.
 This feature is particularly in character with the dashed curve which
 is closer to the one without the form factor.
 The reason is as follows. In the rough, the imaginary potential is
 proportional to Im$\Sigma_s$ $-$ Im$\Sigma_0$ and Im$\Sigma_s$ changes
 from positive sign to negative sign at $E-M$ $\simeq$ $350 MeV$.
   The numerical data of the dipole type
  of form factor with $\Lambda$ = 750 MeV almost correspond to the data
  of the quadrupole type with $\Lambda$ = 1200 MeV. Similarly,
  the dipole type with $\Lambda$ = 1000 MeV corresponds to
  the quadrupole type with  $\Lambda$ = 1500 MeV.
  Our numerical results are a little
 different from empirical information in the
 tendency in the low kinetic energy region.

\centerline{$\underline{\overline{\rm Fig.5(a) , Fig.5(b) }}$}

  The nucleon mean free path is written by using
 Schr$\ddot {\rm o}$dinger equivalent potential form
  as given by Eq.(24).
  In Fig.6 we show the mean free path in the cases of some form
  factors.
  When we choose the dipole type of form factor with
  cut off $\Lambda$=750 MeV
  and the quadrupole type of one with $\Lambda$=1200 MeV
  , the values of mean free
  path are good agreement with the experimental ones, but the experimental
  data are extracted from the reaction of proton and nuclei and so
  we consider that the reasonable form factor is the dipole type with
  $\Lambda$ $\simeq$ 750 $\sim$ 1000 MeV or the quadrupole type with
  $\Lambda$ $\simeq$ 1200 $\sim$ 1500 MeV for nuclear matter.

\centerline{$\underline{\overline{\rm Fig.6 }}$}

  Furthermore,  we calculated the similar calculations using coupling constants
   determined at the normal density
  $\rho$=0.193 ($k_F$=1.42 fm$^{-1}$),
  but the results were nearly equal.

{}~~~~~~~~~~~~~~~~~~~~

{\bf 4. SUMMARY}

{}~~~~~~~~~~~~~~~~~~~~

We evaluated the nucleon mean free path of the energetic nucleon
 based on the BNSD method and compare it with the ones
 extracted experimentally [1], and took a good agreement in the case of
 modification of meson propagators by introducing dipole type
 of form factor
 $\Lambda$ = 750 MeV $\sim$ 1000 MeV and quadrupole type with
  $\Lambda$ = 1200 MeV $\sim$ 1500 MeV. We confirmed that the BNSD
  method was useful for the derivation of optical potential for a
  nucleon traveling far above the Fermi surface
  if we consider the finite size of nucleon or some effects of the
 short-range corrections of nuclear force.

The real self-energy Re$\Sigma^{on}_{SDF}$ grows unphysically
 as the increase of kinetic energy. Even if it can be reasonably
 reduced by modifying meson propagators, the residual Feynman
 self-energy Re$\Sigma^{off}_{SDF}$ is expected also very large.
 We assure again the conclusion discussed in the previous paper [17]
 that we do not pick up Re$\Sigma^{on}_{SDF}$  when we drop out
 Re$\Sigma^{off}_{SDF}$. On the other hand the imaginary part
 of $\Sigma_{SDF}$ is only Im$\Sigma^{on}_{SDF}$.
 The cut off parameter of form factor introduced
 into Im$\Sigma_{SDF}^{on}$
 looks somewhat small but reasonable
 when we regard this parameter as a parameter of nucleon size.

 The real part of optical potential in the
 Schr$\ddot {\rm o}$dinger equivalent form increases
 linearly as the increase of kinetic energy, keeping a good agreement
  with the experimental findings in the lower kinetic energy.

 There remains a problem
 in the imaginary part of optical potential
 that its magnitude is too small
 in the lower kinetic energy region when a form factor is chosen to
 obtain the experimental data of mean free path in the higher kinetic
 energy region.
 In our previous work and this work, we took account of
  the vacuum part of meson self-energy with the low 4-momentum transfer
 to remove a instability around the normal density.
 The vacuum part of meson self-energy is also inserted into the
 denominator of the meson propagator in the integral of
 Im$\Sigma_{SDF}^{on}$. The vacuum part in this case has the high 4-momentum
 transfer and so is expected too much.
  As our next
 work, we will study the role of the vacuum effect for meson
 self-energy in the imaginary potential.

 \centerline{{\bf Acknowledgment }}

The authors are grateful to Prof. T.Kohmura,
 and N.Kakuta for useful discussion, and to the members of nuclear theorist
 group in Kyushyu district in Japan for their continuous encouragement.
 The authors also gratefully acknowledgement the computing time granted by
  Research Center for Nuclear Physics (RCNP).

\vfill\eject

\centerline{{\bf References}}

\noindent
[1]P.U.Renberg, D.F.Measday, P.Pepin, P.Schwaller, B.Favier
     and C.Richard-Serre,

     ~~Nucl.Phys. {\bf A183}(1972)81.

\noindent
[2]B.C.Clark, E.D.Cooper, S.Hama, R.W.Finlay and T.A.Adami,

     ~~Phys. Lett. {\bf B229}(1993)189.

\noindent
[3]E.D.Cooper, S.Hama, B.C.Clark and R.L.Mercer,
     Phys.Rev.{\bf C47}(1993)297.

\noindent
[4]T.Cheon, Phys. Rev. {\bf C38}(1988)1516.

\noindent
[5]R.A.Rego, Phys.Rev. {\bf C44}(1991)1944.

\noindent
[6]G.Q.Li and R.Machleidt, Phys. Rev. {\bf C48}(1993)1062,

\noindent
{}~G.Q.Li, R.Machleidt, R.Fritz, H.M$\ddot {\rm u}$lher and Y.Z.Zhuo,
 Phys. Rev. {\bf 48}(1993)2443.

\noindent
[7]C.Bedau and F.Beck, Nucl. Phys. {\bf A560}(1993)518.

\noindent
[8]X.JI, Phys.Lett. {\bf B208}(1988)19.

\noindent
[9]R.Furnsthl, R.J.Perry and B.D.Serot, Phys. Rev. {\bf C40}(1990)321.

\noindent
[10]K.Lim, Ph.D.Thesis in the Dept.of Phys. Indiana Univ.(1990),

\noindent
{}~K.Lim, and C.J.Horwitz, Nucl. Phys. {\bf A501}(1989)729.

\noindent
[11]M.Prakash, P.J.Ellis and J.I.Kapusta, Phys. Rev. {\bf C45}(1992)2518.

\noindent
[12]J.A.MacNeil, C.E.Price and J.R.Shepard, Phys. Rev. {\bf C47}(1993)1534.

\noindent
[13]M.P.Allends and B.D.Serot, Phys.Rev. {\bf C45}(1992)2975.

\noindent
[14]M.Nakano, A.Hasegawa, H.Kouno and K.Koide,Phys.Rev.{\bf C49}(1994)3076.

\noindent
[15]M.Nakano, K.Koide, T.Mitsumori, M.Muraki, H.Kouno and A.Hasegawa,
  Phys.Rev.

     ~~{\bf C49}(1994)3076.

\noindent
[16]A.Hasegawa, K.Koide, T.Mitsumori, M.Muraki, H.Kouno and M.Nakano,

     ~~Prog.Theor.Phys.{\bf 92}(1994)331.

\noindent
[17]A.Hasegawa, T.Mitsumori, M.Muraki, K.Koide, H.Kouno and M.Nakano,

     ~~Prog.Theor.Phys.{\bf 93}(1995)757.

\noindent
[18]B.Friedman and V.R.Pandharipande,Phys.Lett.{\bf 100B}(1981)205.

\noindent
[19]C.Mahaux and N.Ng$\hat {\rm o}$,Phys.Lett.{\bf 100B}(1981)285.

\noindent
[20]J.D.Walecka,Ann. of Phys.{\bf 83}(1974)491.

\noindent
[21]C,J,Horowitz, Nucl. Phys. {\bf A412}(1984)228.

\vfill\eject

\centerline{{\bf Table and Figure captions}}

\noindent
Table~~~Baryon density (in $fm^{-3}$), coupling constants,
$\sigma$ and $\omega$ meson masses (in MeV) in
        our calculation.

\noindent
Fig.1~~~Feynman diagrams for the calculation of scattering problem
        based on BNSD
        method. Double solid (dotted) lines represent
        exact nucleon(meson)
        propagators, and single solid(dotted)
        lines represent free ones.

\noindent
Fig.2~~~Feynman diagrams for the calculation of the self-energy. (a)Hartree
 diagram, (b)Fock diagram, (c) Higher order diagram more
 than fourth order.
Doubly downward lines denote the hole
        states with momenta under the Fermi momentum $k_F$
	and the doubly
        upward lines denote the intermediate states with momenta $\le k$.

\noindent
Fig.3~~~The real part of nucleon self-energy extracted from
        Feynman propagator of nucleon on-shell. (a) The component
	of scalar part and (b) the component of vector part.

\noindent
Fig.4~~~The imaginary part of nucleon self-energy extracted from
        Feynman propagator of nucleon on-shell.
	(a) The component of vector part and
	(b) the component of scalar part.

\noindent
Fig.5~~~Optical potential in the
        Schr$\ddot {\rm o}$dinger equivalent form. (a)
The imaginary part
         and (b) the real part.

\noindent
Fig.6~~~The nucleon mean free path in nuclear matter. The solid dots with
        error bars
        are experimental results from Ref.[1].

\Large

\centerline{Table}

{}~~~

{}~~~

{}~~~~~

\begin{center}
  \begin{tabular}{ccccc}
     \(\rho_B \)& \(m_s\)&\(m_v\)&\(g_s\) &\(g_v\)    \\ \hline\hline
      0.170&    550  &783    & 9.59   & 11.67       \\    \hline
      0.192&    550  &783    & 9.12   & 11.00       \\    \hline
  \end{tabular}
\end{center}

\end{document}